\documentclass[twocolumn,amsmath,amssymb,floatfix]{revtex4}
\usepackage{graphicx}% Include figure files
\usepackage{bm}% bold math

\newcommand{\beq}{\begin{eqnarray}}
\newcommand{\eeq}{\end{eqnarray}}

\def\der#1#2{\frac{\partial #1}{\partial #2}}

\def \ket#1{|#1\rangle}

\def \ua{{\uparrow}}
\def \da{{\downarrow}}
\def \be{\begin{equation}}
\def \ee{\end{equation}}
\def \ba{\begin{array}}
\def \ea{\end{array}}
\def \bea{\begin{eqnarray}}
\def \eea{\end{eqnarray}}

\def \l{\left}
\def \r{\right}

\def \H{{\cal{H}}}

\def \W{{\Omega}}

\def \D{{\Delta}}
\def \d{{\delta}}

\def \s{s}

\def \f{{\varphi}}
\def \x{{\chi}}

\def \G{{\Gamma}}

\def \av#1{{\overline{\langle#1\rangle}}}

\def \summ{\sum\limits}
\def\nbar{\overline{n}}

\begin{document}

\title{The insulating phases and superfluid-insulator transition of
  disordered boson chains}

\author{Ehud Altman$^1$, Yariv Kafri$^2$, Anatoli Polkovnikov$^3$,
Gil Refael$^4$\\
{$^1$\small \em Department of Condensed Matter Physics, The Weizmann
Institute of Science Rehovot, 76100, Israel}\\
{$^2$\small \em Department of Physics, Technion, Haifa 32000,
Israel}\\
{$^3$\small \em Department of Physics, Boston University, Boston,
MA 02215}\\
{$^4$\small \em Dept. of Physics, California Institute of
Technology, MC 114-36, Pasadena, CA 91125}}

\begin{abstract}

Using a strong disorder real-space renormalization group (RG), we study the phase diagram of a fully disordered chain of interacting
bosons. Since this approach does not suffer from run-away
flows, it allows a direct study of the insulating phases, which are
not accessible in a weak disorder perturbative treatment.  We find that the universal
properties of the insulating phase are determined by the details and symmetries of
the onsite chemical-potential disorder.  Three insulating phases are
possible: (i) an incompressible Mott glass with a finite superfluid
susceptibility, (ii) a random-singlet glass with diverging
compressibility and superfluid susceptibility, (iii) a Bose glass with a finite compressibility but diverging
superfluid susceptibility. In addition to characterizing the
insulating phases, we show that the superfluid-insulator transition is {\em always} of the
Kosterlitz-Thouless universality class.

\end{abstract}

\maketitle

Bose systems can be driven into an insulating phase by
quantum fluctuations due to strong repulsive interactions and lattice effects.
The impact of a disordered potential
on this superfluid-insulator transition and on the nature of the insulating phases is a
long standing question~\cite{FWGF}.
In weakly disordered one dimensional systems, the momentum-shell
renormalization group
(RG) afforded much progress~\cite{Giamarchi-Schulz1987,Giamarchi-Schulz1988}.
But recently, an analysis
using a real space RG suggested that {\it strong} disorder can have
very different effects on a one dimensional Bose system~\cite{AKPR}.
In particular, it was found that a certain type of disorder,
that is perturbatively irrelevant at weak disorder, can actually induce a transition when
sufficiently strong, and lead to a new kind of an insulator termed
the Mott glass \cite{AKPR,LeDoussalGiamarchi1}. The existence of this phase transition and the Mott glass were confirmed numerically~\cite{prokofiev,haas}. The disorder considered in
Refs.~[\onlinecite{AKPR,prokofiev}], however, had a very special
particle-hole symmetry properties that would not be easy to realize in
actual experiments (e.g., \cite{DeMarco,Inguscio}).

\begin{figure}
\includegraphics[width=8.5cm]{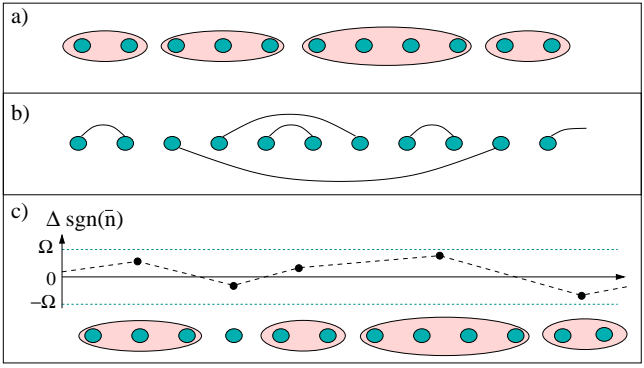}
\caption{The three insulating phases that emerge for
  different classes of disorder. (a) The Mott glass is realized when only the
  Josephson couplings and charging energies are disordered, and with no
  offset charge, $\nbar=0$. At large
  scales it consists of effectively disconnected superfluid
  clusters of random size. (b) The "Random singlet" glass appears when the random
  offset charge is restricted to $\nbar=0,\,1/2$ in terms of
  the basic boson charge. In this phase bosons are delocalized on
  random pairs of remote clusters at all scales. (c) The Bose glass is
  realized for a generic offset-charge distribution. It consists of large
  superfluid clusters acting effectively as weakly coupled
  spin-$1/2$'s in a uniformly-distributed random $z$-field, given by
  the onsite gap times the sign of $\nbar$: $\D sgn(\nbar)= U(1-2|\nbar|)sgn(\nbar)$.}
 \label{flowins}
\end{figure}

In this paper we extend the real-space RG of Ref.~[\onlinecite{AKPR}]
to treat strong and general disorder potentials, not confined to the
comensurability requirement in Ref. \cite{AKPR}.
Our starting point is the disordered quantum-rotor model:
\be
\H=\sum_j {U_j\over 2} \left(\hat{n}_j-\overline{n}_j\r)^2 -\sum_j
J_j \cos\l(\f_{j+1}-\f_j\r), \label{model}
\ee
This Hamiltonian describes a chain of superfluid grains that are connected by a random Josephson
coupling $J_j$ (see Fig. \ref{flowins}(a)). Each grain has a random charging energy $U_j$,
and offset charge $\overline{n}_j$, which represents an excess screening
charge on the site or in its environment. The offset charge parameterizes a
random on-site chemical potential $\mu_j=U_J\overline{n}_j$.
We point out that the lattice model (\ref{model})
can also be derived as a coarse-grained description of continuum bosons, where the grain size
or lattice spacing is set by the healing length of the condensate.

Using a real-space RG, we show that the system can undergo a transition from a superfluid
to three possible insulating phases, whose nature depends on the symmetry properties
of the distribution of offset charges, $\overline{n}_j$.
%The three classes of distributions the integer filling case with $\nbar_j=0$
%everywhere \cite{AKPR}, particle-hole symmetric but incommensurate
%filling where $\nbar_j$ is randomly either $0$ or $1/2$ in each site,
%and the generic case with $-1/2\le \nbar_j<1/2$ with a continuous
%distribution. We describe the insulating phases using three
We characterize the insulating phases using the charging gap $\D$,
the compressibility $\kappa=\der{n}{\mu}$ and the
superfluid susceptibility $\x_s$. The latter corresponds to the linear response of the order parameter
$\av{e^{i\f_j}}$ to the coupling ${\d\over L}\sum_i \cos\f_i$, where
the angular brackets and overline
denote a quantum expectation value and disorder average in that order.
The three insulating phases we find are illustrated schematically in
Fig. \ref{flowins}. They include: (i) an incompressible Mott Glass arising for
the case of zero offset charges $\nbar_j=0$, (ii) a glass phase with a diverging
compressibility which arises if $\nbar_j$ can only take the values of $0$
or $1/2$, and which we term a Random Singlet Glass, and
(iii) a Bose Glass phase characterized by a finite compressibility and
a diverging superfluid
susceptibility in the case of a generic distribution of the offset
charges, i.e., a non-singular distribution in the range
$-1/2<\nbar_j\le 1/2$.
We argue that the superfluid phase and the nature of the
transition are insensitive to the disorder properties of the offset charges.

Our real-space RG analysis, as in random spin chains where it was
first applied ~\cite{MaDas1979,MaDas1980,DSF94}, eliminates the
highest energy scale in the system at each stage through a local decimation step.
The Hamiltonian then maintains its form (Eq. \ref{model}),
but with renormalized distributions of $J_i,U_i$ and $\nbar_i$. From the flow of
the coupling distributions we can learn about both the phases of the
system and its critical points. In particular, the RG provides quantitative predictions
for the properties of the insulating phases and the transition into
them, on which we will concentrate.

We proceed to construct the generalized decimation procedure. Let us define the global energy scale $\Omega=\max_j(\D_j,J_j)$,
where $\D_j=U_j(1-2|\nbar_j|)$ is the charging energy of the site $j$.
For the Hamiltonian in Eq. (1) three types of decimation steps
are possible. {\em Type 1: site decimation}. If $\Omega=\D_j$ for some $j$,
we freeze the charge on the site $j$, thus eliminating this degree of freedom.
A Josephson coupling
$J_{j-1}\approx J_j J_{j+1}/\Omega$ between the sites $j-1$ and $j+1$ is generated
by a virtual tunneling process through the eliminated site (see
Ref.~[\onlinecite{AKPR}]).  {\em Type 2: Bond decimation.} If
$\Omega=J_j$ for some $j$, sites $j$ and $j+1$ merge into a superfluid
cluster with an effective interaction parameter: $1/\tilde
U_{j}=1/U_j+1/U_{j+1}$ (this corresponds to additivity of two
capacitances connected in parallel). The offset charges of the two
sites simply add up $\tilde{\nbar}_j=\nbar_j+\nbar_{j+1}$. {\em Type 3: Doublet formation.}  A special RG step is
introduced for sites with the offset charge $\nbar_j=1/2$. In this
case, if $U_j>\Omega$ the site $j$ is frozen to its two lowest-lying
degenerate charge states. Then we can set $U_j\rightarrow \infty$ and
proceed to treat the site as a spin-1/2 degree of freedom (spin-site)
with $s^z_j=(n_j-\nbar_j)$ and, similarly, $exp(\pm i\phi_j)\rightarrow
\hat{s}^{\pm}_j$. The existence of spin sites requires revisiting
rules for the bond decimation step ({\em Type 2}). If the strong bond
connects a spin site to a regular site, then the spin site can be
simply treated as a half integer site with infinite $U$. But when
two spin-sites are strongly coupled by large $J$ corresponding to
$xy$-coupling: $-J_j(\s^x_js^x_{j+1}+\s^y_js^y_{j+1})$, they freeze
into the triplet state with the total spin $s=1$ and the projection
$m_z=0$:
$\frac{1}{\sqrt{2}}\l(\ket{\uparrow_j\downarrow_{j+1}}+\ket{\uparrow_j\downarrow_{j+1}}\r)$.
This represents a boson resonating in a symmetric superposition
between the two sites, with a "charging" gap  $J_j$ to the excited
(anti-symmetric) state. Quantum fluctuations now allow tunneling
between sites $j-1$ and $j+2$ with strength $J_{j-1}J_{j+1}/J_j$. The
last step is identical to the real-space RG in the random $xx$
spin chain\cite{DSF94}.

For a given system, the RG flow, parameterized by probability
distributions of the couplings, can lead to either a superfluid or
an insulating phase. In the superfluid the system coalesces to one
large superfluid cluster, while in the insulator, it breaks down to
clusters with large effective charging gaps connected by weak
tunneling. Quantitative analysis of the RG flow requires in general
the solution of integro-differential equations for the disorder
distributions~\cite{DSF94}. Remarkably, for the disorder type we
consider, the distributions of $U_i$, $J_i$, and $\nbar_i$, are {\it
universal} in a large vicinity around the fixed points of the RG, which
govern the superfluid-insulator transitions. This
greatly simplifies the flow equations, and in several cases allows an
analytical solution. Below we discuss the three classes of disorder corresponding
to different symmetry properties of the offset-charge
distribution.

{\bf No offset charge.} The case of $\nbar_j=0$, for which the
Hamiltonian (\ref{model}) is particle-hole symmetric, was analyzed
in Ref. [\onlinecite{AKPR}]. For completeness, we review the main results. The RG decimation steps (in this case
only the decimations of {\em Types 1} and {\em 2} are needed) translate
into the flow of the coupling distributions $F(U/\Omega)$ and
$G(J/\Omega)$. Near the superfluid-insulator fixed point they acquire the universal form:
\be
F(x)\approx {A\over x^2}\exp\left[-{f_0(\Omega)\over x}\right],\;
G(x)\approx g_0(\Omega) x^{g_0(\Omega)-1}. \label{dist}
\ee
Here $A$ is a normalization constant, and $x\le 1$~\cite{footnote2}.
Note that the typical strength of the Josephson coupling and site charging
energy monotonically depend on the parameters $g_0$ and $f_0$ respectively.
In [\onlinecite{AKPR}] we derived the following
flow equations for these parameters:
\beq
\frac{df_0}{d\G}=f_0 (1-g_0),\,\,
\frac{dg_0}{d\G}=-f_0, \label{SFeq2}
\eeq
where $\Gamma=\ln(\Omega_0/\Omega)$ and $\Omega_0$ is the initial cutoff energy scale.
The solutions to these equations are parametrized by a single constant $C$: $f_0\approx C+(1-g_0)^2/2$. Negative $C$ corresponds to the superfluid state in which the charging energy is irrelevant ($f_0$ flows to zero) and $g_0$ flows to constant larger than unity. Positive $C$ describes the insulator in which $f_0$ is relevant ($f_0\to\infty$) and
$g_0$, indicative of the typical Josephson coupling strength, flows to
zero. The value $C=0$ corresponds to the critical point separating the two phases. At this point $g_0$ flows to $1$ and $f_0$ flows to zero. We note that substituting $f_0\to y_0^2$ makes Eqs. (\ref{SFeq2}) assume a standard Kosterlitz-Thouless form.

The fixed point in the superfluid phase corresponds to a classical model with $U_i\equiv 0$
and a power-law distribution of the Josephson couplings with an
exponent larger than one. The fact that the fixed point is
noninteracting, implies neither the vanishing of the compressibility,
nor the formation of true long range order: since our analysis relies
on the grand canonical ensemble, the lowest excitation in the superfluid
phase corresponds to an addition of a particle and {\em not}
to a phase twist or to a Bogoliubov excitation. Thus the vanishing of $U_j$ only
implies that the energy for adding an extra particle vanishes with the
inverse system size, as expected in the superfluid phase. Calculation
of the compressibility or stiffness of the {\it superfluid} requires a more detailed analysis,
that keeps track of the internal Josephson couplings and phonon modes within renormalized
clusters.

In this paper we concentrate on the properties of the insulating
phases, which are particularly interesting since they are most
drastically affected by the type of the disorder. In the insulating
phases, the canonical and grand canonical pictures of the excitations are identical, and the real space RG correctly describes their properties.
The insulating phases are best described by a chain consisting of nearly disconnected
clusters, each with its own charging gap. The lowest gap corresponds to the energy
scale $\Omega$ at which the last site is decimated.  In
Ref.~[\onlinecite{AKPR}]  we showed that for the commensurate
case with no offset charges this gap vanishes with system size $L$
as $\Delta\sim \frac{1}{\ln L}$ and is governed by rare and anomalously
large superfluid clusters. The compressibility vanishes
as $(\ln L)/L$. In addition, the insulator is characterized by a
finite superfluid susceptibility. We termed this gapless
and incompressible phase a {\it Mott-glass}. This phase was also discussed
in Refs.~[\onlinecite{LeDoussalGiamarchi1,LeDoussalGiamarchi2}], and confirmed numerically in
Refs.~[\onlinecite{prokofiev, haas}].

%It is a prime example of a Griffiths phase \cite{Griffiths},
%since the gap is not self-averaging (otherwise $\Delta\rightarrow 0$
%would imply $\kappa\rightarrow \infty$) and is dominated by
%anomalously large clusters. {\bf [Question: almost all gapless phases we know
%have (e.g. Metal, Anderson insulator)
%have vanishing gap and finite compressibility,
%so by this definition they are also non self
%averaging, yet they are not Griffiths phases]} - answer: the
%griffiths determination is since the gap in the Mott glass is driven
%to zero by a single point in the system, rather than by a set of
%points which are roughly uniformly distributed.

{\bf Mixed offset: $\bf\nbar=(0,1/2)$}. Let us now allow sites to have either zero or half-integer offset charge maintaining particle-hole symmetry in the problem.
Thus the Hamiltonian (\ref{model}) is invariant under the
transformation $n_j\to 1-n_j$.
Such a restriction naturally arises in an array of small
superconducting grains with a pairing gap much larger
than the charging energy. The random parity of the
electron number in each grain would lead to an offset charge which is randomly either
integer or half-integer in units of the cooper-pair charge.

In addition to distributions of $U$ and $J$, one must now
follow $\nbar$'s distribution as well, which can be
parametrized by the three probabilities --  $q$, $p$, and $s$ -
corresponding to relative densities of integer ($\nbar_j=0$), half
integer ($\nbar_j=1/2$), and doublet (spin-$1/2$) spin-sites
respectively. On first glance, the problem becomes much more
complicated since one has to also consider the flow equations for $p$
and $q$ (note that $p+q+s=1$ due to normalization). Postponing a
detailed description of these equations, let us here observe
that whether the cluster in a renormalized chain is integer or half
integer  depends only on the parity of the number of the bare
sites with $\nbar=1/2$ contained in it. This implies that no matter
what the fraction of the half-integer sites in the original chain was
(as long as it is non-zero) large clusters  have odd or
even parity with equal probabilities. For this
reason the distribution quickly flows to a fixed line with
$p=q=(1-s)/2$, provided $p>0$ initially.

This observation significantly simplifies the flow analysis, and
reduces the number of flow equations to three. A detailed analysis of
the integro-differential equations shows that the distributions of
 $U$ and $J$ flow to the same universal distributions as
those described by Eqs.~(\ref{dist}), and we obtain a simplified set of flow
equations:
\beq
\frac{d f_0}{d\G}&=&f_0\l[1-g_0(1-\s)(1+f_0)\r],\nonumber\\
\frac{d g_0}{d\G}&=&-{g_0\over 2}\l[(1-\s)f_0+2\s^2 g_0\r],\label{redeq}\\
\frac{d\s}{d\G}&=&{f_0\over 2}(1-\s^2)-g_0\s(1-\s).\nonumber
\eeq
These have two different families of solutions, corresponding to
two different phases. In one family, the flow is to a stable
fixed line with $s=0$, $f_0=0$ and $g_0>1$: a superfluid phase identical to that of the no offset charge case. Close to this line, the flow equations reduce to those for the integer
case Eqs.~(\ref{SFeq2}), except for an extra factor of $1/2$ in the
equation for $g_0$. This factor enters because only integer sites with
a charge gap can renormalize $J$. The fact that the random offset is
not important for describing the superfluid is not surprising:
large local particle number fluctuations in this phase completely
overwhelm small fluctuations in the offset charge.

The superfluid-insulator transition is also very similar to the
$\nbar=0$ case, and it is described by essentially the same
Kosterlitz-Thouless critical
point with $g_0=1$, $f_0=0$ and $s=0$, similar to
Eqs. (\ref{SFeq2}). In the insulating phase at
$g_0<1$, however, $f_0$ becomes relevant, as well as the spin-site density $s$, which quickly flows to
$s=1$. This second family of solutions of (\ref{redeq}) describes a completely different insulating phase than in
the case of zero offset charges; it corresponds to an {\em effective
  spin-$1/2$ chain with random $x-y$ ferromagnetic couplings}, a
model that was analyzed in detail in Ref.~[\onlinecite{DSF94}].
Its ground state consists of random non-crossing pairs of sites at varying distances,
in which the spins form the $m_z=0$ state, $\frac{1}{\sqrt{2}}\l(\ket{\ua\da}+\ket{\da\ua}\r)$.
This phase is termed the Random-Singlet Glass, reafirming the
connection with anti-ferromagnetic random spin chains~\cite{BhattLee,FisherDoty}.
In the bosonic language the ground state has bosons delocalized
randomly between pairs of sites.

Many properties of the Random-Singlet Glass can be found directly from the
analysis of Ref.~[\onlinecite{DSF94}]. The energy scale associated with breaking a
singlet between sites of distance $\ell$ is
$\Omega_{\ell}=\W_0\exp(-\sqrt{\ell})$. By setting $\ell=L$, the system size,
we obtain the scaling of the gap vs. the system size. Following the identification $n_i=\frac{1}{2}+\hat{s}_i^z$, the compressibility $\kappa$ and superfluid susceptibility $\chi_s$ in the insulating phase
correspond to the susceptibilities of a random spin chain to a Zeeman field applied in the $z$ and $x$ directions respectively. From Ref. [\onlinecite{DSF94}] we see that both susceptibilities diverge at the limit of small $\Omega$ as $\kappa(\Omega)\sim \chi_s(\Omega)\sim 1/\Omega\log^3(\W_0/\Omega)$.
While the superfluid stiffness vanishes in the
thermodynamic limit, unlike the stiffness of the Mott glass, it vanishes
only sub-exponentially with system size: $\rho_s\propto
e^{-\sqrt{L}}$. In addition, note that $g_0$, indicative of 
the strength of the Josephson coupling, flows to zero as $g_0\sim
1/\Gamma$ (as results from Eq.~(\ref{redeq}) with $s=1$), much slower
than $g_0\sim \exp[e^{-\Gamma}]=\exp[-1/\Omega]$ in the Mott glass.

{\bf Generic chemical potential disorder.}  When all possible offsets
$-1/2<\overline n_j\le 1/2$ are allowed, the relevant energy scale for the
RG site decimation is not $U$, but rather the local gap
$\D_i=U_i(1-2|\nbar_i|)$. The interaction
$U_i$ is allowed to exceed $\Omega$ by an amount which depends on
the local charge offset: 
\be
U_j< \Omega/(1-2|\nbar_j|).
\label{maxU}
\ee
Thus we
must consider a {\em joint} distribution for $U$ and $\nbar$, which
makes the RG flow equations rather cumbersome. Their analysis, however,
reveals a rather intuitive behavior. Below we describe the key
steps of the derivation. 

The first step is to note that
$\nbar$ disorder width is a relevant variable due to the rule
$\overline n_j\to \overline n_j+\overline n_{j+1}$ under the
decimation of a strong bond. This implies that as the effective sites
grow with the RG, their offset charges quickly become uniformly
distributed between $-1/2$ and $1/2$, i.e., the largest disorder allowed. This observation significantly
simplifies the analysis. In particular, it is straightforward to check
that the distributions of $U$ and $J$ again approach the universal
functions~(\ref{dist}). In the superfluid regime and at the transition
point the flow is also governed by the equations~(\ref{SFeq2}) with
an extra factor of one half in the equation for $g_0$. Thus the system
with generic disorder undergoes the same transition as in the two
cases discussed above. Similarly, the elementary understanding of this
result is that for small interactions the bounded disorder in the
offset charge is overwhelmed by large particle number fluctuations. In
the insulating side, $f_0$ flows to large values; the joint
distribution of $U/\Omega$ and $\nbar$ is still given by (\ref{dist}):
$F^{joint}(U/\Omega,\nbar)=F(U/\Omega)$, but with the upper bound of
$U/\Omega$ obeying Eq. (\ref{maxU}). At large $f_0$, this form of $F^{joint}(U/\Omega,\nbar)$ leads to the uniform distribution of
the charging gaps $\D$ of each site: $H(\D)\approx 1/\Omega$. Most
importantly, $H(\D)$ is non-singular at $\Delta\rightarrow 0$.

Deep in the insulating phase, since the typical charging energy $U$ is
large, each site can be treated as a doublet of states which is split
locally by the energy $\D<\Omega$, while all other states lie at energies
above the cutoff. When $\Delta=0$ this doublet is exactly degenerate
and represents the spin-$1/2$ degree of freedom discussed above. Thus the spin-$1/2$ description is also applicable to the case of
nonzero $\Delta$ with the latter playing the role of the Zeeman field
along the $z$-axis. The effective Hamiltonian describing the chain
becomes
\be
\H=-\summ_j
\tilde{J}_j\l(s^x_js^x_{j+1}+s^y_js^y_{j+1}\r)-\summ_j
  \D_j s^z_j,
\ee
where $s^z_i=n_i-1/2$, and here
$\Delta_j=U_j(1-2|\nbar_j|)sgn(\nbar_j)$ can be both positive and
negative (its distribution is $H(\D sgn(\nbar))=1/2\Omega$).
Because $\tilde{J}_j\ll\Omega$ we can
calculate many properties of this phase by considering a single site
subject to a random Zeeman field. First, as implied by the gap distribution,
in this phase the energy-length scaling is $\Omega_L\sim 1/L=\rho$. Next, the compressibility is
given by the response to an external field in the $z$ direction:
$\kappa=\rho\partial \av{s^z}/\partial h^z_{ext}=2
H(0)=\rho/\Omega=\kappa_0$, a constant at low energies. The
superfluid susceptibility is the response to a transverse field:
$\x_s=\rho\partial\av{s^x}/\partial h^x_{ext}$. We find this disorder average from the distribution of $\Delta_i$:
\be
\chi_s=\rho\int_{-\Omega}^\Omega {d\Delta\over
2\Omega}{1\over\sqrt{(h^x_{ext})^2+\Delta^2}} \approx {\kappa_0\over 2}
log\left({\Omega\over |h^x_{ext}|}\right),
\ee
which diverges as $h^x_{\rm ext}\to 0$, with a functional form resulting from the non-singular behavior of $H(\Delta)$. In a finite chain this divergence is cut off
by the smallest $|\Delta_j|$, and leads to $\chi_s\sim \log L$. These properties, namely finite
compressibility and diverging superfluid susceptibility, coincide with
the Bose Glass phase discussed in Ref. ~[\onlinecite{FWGF}].

\begin{table}[ht]
\begin{center}
\begin{tabular}{|c|c|c|c|c|}
\hline $\nbar$ disorder & Glass type & $\Delta$ & $\kappa$ & $\chi_{s}$ \\
\hline $0$ & Mott & $\frac{1}{\log L}$ & $\frac{\log L}{L}\rightarrow 0$
& const \\
$0,\,1/2$ & $\ba{c}\mbox{Random-}\\ \mbox{singlet}\ea$ & $e^{-c\sqrt{L}}$ &
$\frac{1}{L^{3/2}}e^{c\sqrt{L}}$ & $\frac{1}{L^{3/2}}e^{c\sqrt{L}}$ \\
$-1/2\le\nbar<1/2$ & $\ba{c} \mbox{}\\ \mbox{} \ea$Bose & $\frac{1}{L}$ & $\kappa$ & $\frac{\kappa}{2}\log L $\\
\hline

\end{tabular}
\caption{The gap $\Delta$,
  compressiblity $\kappa$, and SF susceptibility $\chi_s$, of
  the insulating phases realized for the different classes of disorder
  in the offset charge $\bar n$. Here $L$ is the system size and $c$
  denotes a nonuniversal constant.   
\label{instable}
}
\end{center}
\end{table}

In summary, we extended the real-space RG analysis of Ref.
[\onlinecite{AKPR}] to address non-commensurate random chemical
potential. We found that the symmetry of the Hamiltonian and the
details of the disorder, as encoded in the distribution of the
offset $\nbar_j$, do not
affect the universal properties of the superfluid phase and
superfluid-insulator transition, but at the
same time it completely determines the insulating phase of the
system. In addition to the Mott-Glass of the
integer filling case, we find the Random-Singlet Glass for the mixed
$\nbar=0,1/2$ case, and for generic chemical-potential disorder, a
phase we identify as the Bose-Glass\cite{FWGF}. We note that if we restrict
$\nbar$ to a set of rational numbers, we would still obtain either the
random-singlet glass, or the Mott-glass. All our results
are supported by a numerical real-space RG study to be published separately.

Our focus here was the universal thermodynamic properties of the
insulating phases, as summarized in Table \ref{instable}. An outstanding question, which we leave to a future publication, is the effective
Luttinger parameter at the critical point, and how it compares to the
low-disorder motivated results of Ref.~[\onlinecite{Giamarchi-Schulz1987}],
predicting $K=\sqrt{\kappa\rho_s}=3/2$. Our initial results, however, indicate that
the critical value of the Luttinger parameter at sufficiently strong disorder is
non-universal, and exceeds $3/2$; this possibility does not contradict a general thermodynamic statement of
Ref.~[\onlinecite{prokofiev}].

{\em Acknowledgments.} We thank S.~Girvin for suggesting
to consider the half integer case. We also acknowledge T.
Giamarchi, P. Le-Doussal, O. Motrunich, N. Prokofe'v, and B. Svistunov
for numerous discussions. A.P.'s research was supported by AFOSR
YIP, Y.K. and G.R. acknowledge Boston University visitors program's
hospitality. Y.K. and E.A. thank the ISF and BSF for support.

\bibliography{random}

\begin{thebibliography}{16}
\bibitem[{foo()}]{footnote2}
\bibinfo{howpublished}{For the commensurate case $A=e^{f_0}/f_0$, but for the
  generic disorder case, where $0<x<\infty$, it will have a different value.}
\end{thebibliography}

\end{document}